\documentclass[prb,twocolumn,amsmath,amssymb,showpacs]{revtex4}
\usepackage{graphicx}
\newcommand{\beq}{\begin{equation}}
\newcommand{\eeq}{\end{equation}}
\begin{document}
%\tightenlines
%\begin{normalsize}

\title{Collective excitations in unconventional charge-density
wave systems}
\author{{A. Greco}$^{a,b}$ and {R. Zeyher}$^a$ } 
\affiliation{$^a$Max-Planck-Institut\  f\"ur\
Festk\"orperforschung,\\ Heisenbergstr.1, 70569 Stuttgart, Germany \\
$^b$Permanent address: Departamento de F\'{\i}sica, Facultad de
Ciencias Exactas e Ingenier\'{\i}a and \\
IFIR(UNR-CONICET), Av. Pellegrini 250, 
2000-Rosario, Argentina}
\date{\today}

\begin{abstract} 
The excitation spectrum of the $t$-$J$ model is studied on a square lattice in
the large $N$ limit in a doping range where a $d$-$density$-$wave$ (DDW)
forms below a transition
temperature $T^\star$. Characteristic features of the DDW ground state
are circulating currents which fluctuate above and condense into a staggered 
flux state below $T^\star$ and density fluctuations where the electron and the
hole are localized at different sites. General expressions for the density
response are given both above and below $T^\star$ and applied to Raman,
X-ray, and neutron scattering. Numerical results show that the density response
is mainly collective in nature consisting of broad, dispersive structures 
which transform into well-defined peaks mainly at small momentum transfers. 
One way to detect these excitations is by inelastic neutron 
scattering at small momentum transfers where the cross section 
(typically a few per cents of that for spin scattering) 
is substantially enhanced, exhibits a strong dependence on the 
direction of the transferred momentum and a well-pronounced peak somewhat
below twice the DDW gap. Scattering from the DDW-induced Bragg peak is found
to be weaker by two orders of magnitude compared with the momentum-integrated
inelastic part. 
\end{abstract}
\maketitle

%\begin{multicols}{2}
%\newpage
\section{Introduction}

The electron density operator $\rho$ on a lattice has in momentum
space the general form,
\begin{equation}
\rho({\bf q}) = {1\over N_c}\sum_{{\bf k} \sigma} \gamma({\bf k},{\bf q})
c^\dagger_\sigma ({\bf k + q}) c_\sigma (\bf k).
\label{rho}
\end{equation} 
$c^\dagger_\sigma({\bf k}), c_\sigma(\bf{k})$ are electron creation
and annihilation operators with momentum $\bf k$ and spin projection
$\sigma$, $N_c$ is the number of primitive cells, 
$\bf q$ the transferred momentum, and $\bf k$ 
characterizes the relative motion of
electron and hole. If both particles reside always on the same lattice site 
$\gamma$ is independent on $\bf k$ and $\rho$ describes a usual density
wave. This wave may condense for some wave vector $\bf q$ different
from the reciprocal lattice vectors of the unmodulated lattice and a
conventional charge density wave (CDW) state is obtained. If the 
electron and hole explore different sites $\gamma$ depends on 
$\bf k$ and the condensation of this wave
yields a CDW state with an internal symmetry described by the ${\bf k}$
dependence of $\gamma$.
We will call 
this state in the following an unconventional CDW state\cite{Maki1}. 
Proposals for systems
with an unconventional CDW state include 
organic conductors\cite{Maki1,Maki2} and the pseudogap phase of 
high-temperature superconductors\cite{Hsu,Cappelluti,Benfatto,Chakravarty}. 

Static and fluctuating density waves of the general form of Eq.(\ref{rho})
can be observed by various experimental probes. The vector potential
generated by the magnetic moment of neutrons influences the hopping
elements of electrons via the Peierls substitution. Thus neutrons probe
orbital magnetic fluctuations associated with density fluctuations.
As shown in Ref.\onlinecite{Hsu} the resulting neutron cross section
is related to the imaginary part of a retarded 
density-density correlation function, where $\gamma$ is given by,
\begin{eqnarray}
\gamma({\bf k},{\bf q}) &=& {{8i\pi e \mu_0 t_{eff}}\over{\hbar c |{\bf q}|}} 
\times \nonumber \\
\Bigl( \sum_{j=x,y} {\alpha_j\over q_j} &(&(1-\cos{q_j})\cos{k_j}
 +\sin{q_j} \sin{k_j})
\Bigr),
\label{gamma}
\end{eqnarray}
with
\boldmath
\begin{equation}
{\alpha} = {\mu \over{\mbox{\unboldmath$\mu_0$}}}\times 
{q\over|q|}\mbox{\unboldmath$.$}
\label{alpha}
\end{equation}
\unboldmath

The expression Eq.(\ref{gamma}) applies to a solid of square layers
with lattice constant $a$ which we put, together with $\hbar$ to one in the 
following.
$t_{eff}$ is an effective nearest neighbor hopping integral,
$c$ the velocity of light, \boldmath ${\mu}$ \unboldmath 
the magnetic moment of the
neutrons, and 
$\mu_0 = \sqrt{{{\mbox{\boldmath$\mu$}\unboldmath ^2}\over 3}}$. 
We neglect hopping between the layers as well as 
a small contribution to $\gamma$ from second-nearest neighbor
hoppings $\sim t'$.
Similarly, inelastic light and non-resonant X-ray scattering
are determined by the Raman or the X-ray scattering
amplitude which corresponds to the following expression for 
$\gamma$\cite{Devereaux},
\begin{equation}
\gamma({\bf k},{\bf q}) = \sum_{\alpha \beta} e^s_\alpha {{\partial^2 
\epsilon ({\bf k}+{\bf q}/2)}\over{\partial k_\alpha \partial k_\beta}}
e^i_\beta.
\label{Raman}
\end{equation}
${\bf e}^i$ and ${\bf e}^s$ are the polarization vectors of the
incident and scattered light, respectively, and 
$\epsilon({\bf k})$ the one-particle energy. In contrast to 
Refs.\onlinecite{Hsu,Devereaux} we assumed that the Peierls substitution
is made in the renormalized Hamiltonian so that the renormalized
hopping $t_{eff}$ and one-electron band appear in 
Eqs.(\ref{gamma}) and (\ref{Raman}).
For Raman scattering the momentum transfer is practically zero
whereas no such restriction exists for X-ray scattering. 

Microscopic models for interacting electrons usually contain interactions
between local charge densities or spin densities, such as, the Coulomb or 
Heisenberg interaction. $\bf k$-dependent densities of the form of
Eq.(\ref{rho}) become important if the exchange terms are 
competing or dominating the direct, Hartree-like terms. The excitonic
insulator is such a case where the Coulombic exchange terms not only
create excitons but force them to condense into a new ground 
state\cite{Rice}. These ideas were applied to high-T$_c$ cuprates by 
Efetov\cite{Efetov} assuming that the barely screened Coulomb interaction
between layers produces interlayer excitons and their condensation.
More recently, it was recognized that the $t$-$J$ model
in the large $N$ limit ($N$ is the number of spin components) shows 
a transition to a charge-density wave state with internal
$d$-wave symmetry (DDW) at low doping\cite{Cappelluti} with a transition 
temperature $T^\star$.
This model describes electrons in the $CuO_2$ planes of high-T$_c$ cuprates
and identifies the pseudogap phase at low dopings with a DDW 
state. One may expect that near the phase boundary to the DDW  
state or in the DDW state density fluctuations of the form of Eq.(\ref{rho})
become important and, according to the above discussion, could be
detected by neutron or X-ray scattering. 
It is therefore the purpose of this paper to calculate the general 
density response of this model, both above and below $T^\star$, 
and to make predictions for the
magnitude and the momentum and polarization dependence of neutron and 
X-ray cross sections.    

The order parameter of a commensurate DDW state with wave vector $(\pi,\pi)$
is purely imaginary which follows from the hermiticity of the large $N$
Hamiltonian. As a consequence, circulating, local currents accompagny
in general density fluctuations in the DDW state producing
orbital magnetic moments localized on the plaquettes of the square lattice.
The circulating currents fluctuate above and freeze into
a staggered flux phase below $T^\star$. 
Most previous calculations considered ${\bf q}=0$ quantities such as
the frequency-dependent conductivity\cite{Benfatto1,Aristov,Benfatto2,Carbotte}
or Raman scattering\cite{Zeyher4,Carbotte}. Calculations for finite
momentum transfers\cite{Tewari} to be presented below are interesting 
because they probe
the local properties of the density fluctuations and the flux lattice.

In section II we will first reformulate the large $N$ limit of the 
$t$-$J$ model
in terms of an effective Hamiltonian which contains usual
electron creation and annihilation operators (i.e., no constraint
operators), renormalized bands and an interaction between several charge
density waves. This effective model contains as a special case 
the models used in Refs.\onlinecite{Maki1,Carbotte} in discussing 
anomalous charge density waves. 
We then extend previous normal-state calculations of the density response, 
given by a 6x6 susceptibility matrix $\chi$, to the DDW state using the
Nambu formalism. The obtained expressions hold for all momenta $\bf q$ and 
are general enough to discuss the dispersion of collective modes and
the momentum and polarization dependencies of experimental cross sections.

Section III contains numerical results for the formulas presented in
section II. In subsection A the density fluctuation spectrum
in the d-wave channel will be discussed as a function of momentum, both
above and below $T^\ast$. In subsections B and C results for the cross 
sections for non-resonant inelastic X-ray scattering and for polarized and
unpolarized neutron scattering will be given, 
and subsection D compares the magnitude of $\bf q$ integrated neutron cross 
sections for spin and for orbital scattering. Section IV contains our
conclusions.   

\section{Density response in the d-CDW state}

In Refs.\onlinecite{Zeyher1,Zeyher2} it has been shown that 
the density response in the normal state of the $t$-$J$ model at large $N$
can be obtained by using the following two sets of density operators,
\begin{equation}
\rho^{\prime}_\alpha({\bf q}) = {1\over N_c} \sum_{{\bf k}\sigma}
E_\alpha({\bf k},{\bf q}) c^\dagger_\sigma({\bf k}+{\bf q})c_\sigma({\bf k}),
\label{EF1}
\end{equation}
\begin{equation}
\rho_\beta({\bf q}) = {1\over N_c} \sum_{{\bf k}\sigma}
F_\beta({\bf k})c^\dagger_\sigma({\bf k}+{\bf q}) c_\sigma({\bf k}),
\label{EF2}
\end{equation}
with
\begin{equation}
E_\alpha({\bf k},{\bf q}) =(1,t({\bf k}+{\bf q})+J({\bf q}),
\gamma_3({\bf k}),\gamma_4({\bf k}),\gamma_5({\bf k}),\gamma_6({\bf k})),
\end{equation}
and
\begin{equation}
F_\beta({\bf k}) = (t({\bf k}),1,2J\gamma_3({\bf k}),
2J\gamma_4({\bf k}),2J\gamma_5({\bf k}),2J\gamma_6({\bf k})).
\end{equation}
Compared to Refs.\onlinecite{Zeyher1,Zeyher2} a slight change in the 
representation 
of the basis functions has
been made by using the symmetrized functions
\begin{eqnarray}
\gamma_{3,5}& =& (\cos{k_x} \pm \cos{k_y})/2, \nonumber \\
\gamma_{4,6}& =& (\sin{k_x} \pm \sin{k_y})/2,
\label{basis}
\end{eqnarray}
where the subscripts 3 and 4 refer to the $+$ and 5 and 6 to the $-$ sign.
Here, and in the following, we put the lattice constant of 
the square lattice to $1$.  
$t({\bf k})$ and $J({\bf k})$
are the Fourier transforms of the hopping amplitudes $t_{ij}$ and
the Heisenberg term $J_{ij}$, respectively.
Including nearest and second nearest neighbor hoppings $t$ and $t'$, 
respectively, we have 
\begin{equation}
t({\bf k}) = 2t(\cos{k_x} + \cos{k_y}) +4t'\cos{k_x} \cos{k_y}.
\label{dispersion}
\end{equation}
The corresponding density-density Matsubara Green's function matrix
is defined by
\begin{equation}
\chi_{\alpha\beta}({\bf q},i\omega_n) = -\int_0^{1/T} d\tau e^{-i\omega_n\tau}
<T_\tau\rho'_\alpha({\bf q}\tau)\rho^\dagger_\beta({\bf q}0)>.
\end{equation}

The general solution for $\chi$ in the normal
state has been given in the large $N$ limit and the dispersion of
macroscopic density fluctuations, determined  by the element $\chi_{12}$,
have been discussed\cite{Gehlhoff,Zeyher1}. 
At lower temperatures and dopings the density
component $E_5({\bf k},{\bf Q})$ freezes in for a wave vector $\bf Q$
approximately equal to $(\pi,\pi)$, leading to the DDW state.

The Hilbert space of the $t$-$J$ model does not contain double
occupancies of sites, its operators are therefore $X$ 
and not the usual creation and annihilation operators $c^\dagger_{i\sigma}$
and $c_{i\sigma}$ of second quantization. In the large $N$ limit
the $t$-$J$ model becomes, however, equivalent to the following 
effective Hamiltonian in terms of usual creation and annihilation 
operators,
\begin{equation}
H_{eff} = \sum_{{\bf k}\sigma} \epsilon({\bf k}) c^\dagger_\sigma({\bf k})
c_\sigma({\bf k}) - {N_c\over 2}\sum_{\alpha=1}^6 \sum_{\bf q}
\rho_\alpha^{\prime}({\bf q})\rho_\alpha^\dagger({\bf q}).
\label{Heff}
\end{equation}
$\epsilon({\bf k})$ are the one-particle energies in the large $N$ limit,
\begin{equation}
\epsilon({\bf k}) = {\delta \over 2}t({\bf k}) - {J({\bf k})\over 2} \cdot
{1\over N_c}\sum_{\bf p} \cos(p_x)f(\epsilon({\bf p})-\mu).
\label{eps}
\end{equation}
$\mu$ is the renormalized chemical potential and $f$ the Fermi function. 
The second term in Eq.(\ref{Heff})
represents an effective interaction originating from the functional
derivative of the self-energy with respect to the Green's function.
This interaction is hermitean though its present form does not show this
property explicitly because of the chosen compact form to represent it.
Using Eqs.(\ref{EF1}) and (\ref{EF2}) one easily verifies that the sum of the 
terms 
$\alpha=3...6$ represent the Heisenberg interaction, written as a charge-
charge interaction and as a sum of four separable kernels. The terms
$\alpha=1,2$ originate from the constraint which acts in $H_{eff}$ as a
two-particle interaction. $H_{eff}$ and the original $t$-$J$ Hamiltonian in the
large $N$ limit become equivalent if only bubble diagrams generated by the
second term in $H_{eff}$ and no self-energy corrections are taken into account.
In the DDW state $<\rho_\alpha({\bf Q})>$  and $<\rho'_\alpha({\bf Q})>$
become non-zero for $\alpha=5$.
We rewrite $H_{eff}$ as,
\begin{eqnarray}
H_{eff} &=& \sum_{{\bf k}\sigma} \epsilon({\bf k}) c^\dagger_\sigma({\bf k})
c_\sigma({\bf k}) - N_c<\rho'_5({\bf Q})> \rho^\dagger_5({\bf Q})\nonumber \\  
&-&{N_c\over 2}\sum_{\alpha=1}^6 \sum_{\bf q}
\tilde{\rho}_\alpha^{\prime}({\bf q})\tilde{\rho}_\alpha^\dagger({\bf q}).
\label{Heff1}
\end{eqnarray}
The tilde at the density operators means that only their fluctuating
parts should be considered. 

In order to be able to use usual diagrammatic rules in the DDW state
we introduce the Nambu notation, i.e., the row vector 
$\psi^\dagger_\sigma({\bf k}) = (c^\dagger_\sigma({\bf k}),
c^\dagger_\sigma({\bf k}+{\bf Q}))$ and the corresponding column
vector obtained by hermitean conjugation. We also make the convention
that the momentum in $\psi$ or $\psi^\dagger$ are always taken modulo
the reduced Brillouin zone (RBZ), i.e., lie in the RBZ. The momentum sums
over the (large) Brillouin zone (BZ) are then split into parts inside
and outside of the RBZ and the two parts then combined using the Nambu
vectors and Pauli matrices. We obtain then,
\begin{equation}
\rho'_\alpha({\bf q}) = {1\over N_c} {\sum_{{\bf k}\sigma}}^\prime
\psi^\dagger_\sigma({\bf k}+{\bf q})\hat{E}_\alpha({\bf k},{\bf q})
\psi_\sigma({\bf k}),
\label{rhoalpha}
\end{equation}
\begin{equation}
\rho_\beta({\bf q}) = {1\over N_c} {\sum_{{\bf k}\sigma}}^\prime
\psi^\dagger_\sigma({\bf k}+{\bf q})\hat{F}_\beta({\bf k})
\psi_\sigma({\bf k}),
\label{rhobeta}
\end{equation}
with
\begin{eqnarray}
\hat{F}_1({\bf k})&=& t^{(1)}({\bf k}) P_o({\bf k}+{\bf q},{\bf k})
+t^{(2)}({\bf k}) P_e({\bf k}+{\bf q},{\bf k}), \nonumber \\
\hat{F}_2({\bf k})&=& P_e({\bf k}+{\bf q},{\bf k}), \nonumber \\
\hat{F}_\beta({\bf q}) &=&  F_\beta({\bf k}) P_o({\bf k}+{\bf q},{\bf k}),
\label{F}
\end{eqnarray}
for $\beta=3...6$,
\begin{eqnarray}
\hat{E}_1({\bf k},{\bf q}) &=& \hat{F}_2({\bf k}), \nonumber \\
\hat{E}_2({\bf k},{\bf q})&=& t^{(1)}({\bf k}+{\bf q}) 
P_o({\bf k}+{\bf q},{\bf k})\nonumber\\ 
&+&(t^{(2)}({\bf k}+{\bf q})+J({\bf q})) P_e({\bf k}+{\bf q},{\bf k}), 
\nonumber \\
\hat{E}_\alpha({\bf k},{\bf q}) &=& \hat{F}_\alpha({\bf k})/2J,
\label{E}
\end{eqnarray}
for $\alpha=3...6$,
and
\begin{eqnarray}
P_o({\bf k}+{\bf q},{\bf k})&=&p({\bf k}+{\bf q},{\bf k})\sigma_3
-i(1-p({\bf k}+{\bf q},{\bf k}))\sigma_2, \nonumber \\
P_e({\bf k}+{\bf q},{\bf k})&=&p({\bf k}+{\bf q},{\bf k})\sigma_0
+(1-p({\bf k}+{\bf q},{\bf k}))\sigma_1. \nonumber \\
\end{eqnarray}
$t^{(1)}({\bf k})$ and $t^{(2)}({\bf k})$ are the Fourier transforms
of the nearest and second-nearest neighbor hopping amplitudes,
$\sigma_0$ is the 2x2 unit matrix, and $\sigma_1,\sigma_2,\sigma_3$
are Pauli matrices. The dash at the summation sign in Eq.(\ref{rhobeta})
means a restriction of the sum to the RBZ. $p({\bf k}+{\bf q},{\bf k})$
with ${\bf k}$ in the RBZ is equal to one if ${\bf k}+{\bf q}$,
reduced to the BZ, lies in the RBZ and is zero otherwise.
Finally, the first and 
second terms in $H_{eff}$ can be combined and read in Nambu notation as
\begin{eqnarray}
H^{(0)}_{eff} = {\sum_{{\bf k}\sigma}}^\prime \psi^\dagger_\sigma({\bf k})
\Big[(\epsilon_+({\bf k})-\mu)\sigma_0 \nonumber \\
+\epsilon_-({\bf k})\sigma_3
+\Phi \gamma_5({\bf k}) \sigma_2\Big]
\psi_\sigma({\bf k}).
\label{H0}
\end{eqnarray}
The energies $\epsilon_\pm$ are defined by $(\epsilon({\bf k}) \pm
\epsilon({\bf k}+{\bf Q}))/2$.
$\Phi$ is an abbreviation for the amplitude of the DDW and must be
real because of the hermiticity of $H_{eff}$.

Using the above Nambu representation the susceptibility matrix
$\chi_{\alpha \beta}$ is obtained by a bubble summation, 
\begin{equation}
\chi_{\alpha\beta}(q) = \sum_{\gamma} 
\chi^{(0)}_{\alpha\gamma}(q)
(1+\chi^{(0)}(q))^{-1}_{\gamma\beta}.
\label{chi}
\end{equation}
Here we used the abbreviation $q=(i\nu_n,{\bf q})$ with the bosonic
Matsubara frequencies $\nu_n = 2\pi T n$, where $T$ is the temperature.
$\chi^{(0)}(q)$ stands for a single bubble and is given analytically
by the expression
\begin{equation}
\chi_{\alpha \beta}^{(0)}(q) = {T\over N_c} {\sum_{{\bf k}n}}^\prime
Tr\Big[ G(k+q)\hat{E}_\alpha({\bf k},{\bf q})G(k)\hat{F}^\dagger_\beta({\bf k})
\Big],
\label{chi0}
\end{equation}
with the Nambu Green's function
\begin{equation}
G(k) = {{(i\omega_n+\mu-\epsilon_+({\bf k}))\sigma_0 
+\epsilon_-({\bf k})\sigma_3 
+\Phi \gamma_5({\bf k})\sigma_2}\over{(i\omega_n -E_1({\bf k}))(i\omega_n-
E_2({\bf k}))}},
\label{G}
\end{equation} 
and the two eigenenergies 
\begin{equation}
E_{1,2}({\bf k}) = \epsilon_+({\bf k}) -\mu \pm \sqrt{\epsilon_-^2({\bf k})
+\gamma_5^2({\bf k}) \Phi^2}.
\end{equation}
After performing the trace over Pauli matrices and the sum over fermionic 
Matsubara frequencies in Eq.(\ref{chi0}) 
only the sum over momenta is left for a numerical evaluation.
In the normal state we obtain
\begin{eqnarray}
& &\chi_{\alpha \beta}^{(0)}({\bf q},i\omega_n) = \nonumber\\ 
& &{1\over N_c} \sum_{\bf k}
E_\alpha ({\bf k},{\bf q})F_\beta ({\bf k}) {{f(\epsilon({\bf k}+{\bf q}))
-f(\epsilon({\bf k}))}\over {\epsilon({\bf k}+{\bf q}) -\epsilon({\bf k})
-i\omega_n}}.
\label{chi00}
\end{eqnarray}

Two special cases of Eq.(\ref{chi}) should be mentioned. If $\bf q$
is parallel to the diagonal, i.e., of the form ${\bf q} = (q,q)$,
the components 5 and 6 in the matrices, which are odd under reflection
at the diagonal, decouple from the other components. At the points
${\bf q}=(0,0)$ and $(\pi,\pi)$ they even decouple
from each other so that Eq.(\ref{chi}) can be solved by division,
for instance\cite{Zeyher3},
\begin{equation}
\chi_{55}(q) = \chi_{55}^{(0)}(q)/(1+\chi_{55}^{(0)}(q)),
\label{chi55}
\end{equation}
for ${\bf q}= (0,0)$ and ${\bf q}=(\pi,\pi)$. In the first case
$\chi_{55}$ describes the $B_{1g}$ component of Raman scattering\cite{Zeyher4},
in the second case amplitude fluctuations of the order parameter\cite{Tewari}.
The other special case refers to a DDW without 
constraint\cite{Tewari,Carbotte}.
Since the components 1 and 2 were caused by the constraint they can
be dropped in this case and Eq.(\ref{chi}) reduces to a 4x4 matrix
equation. The four components arise because a nearest neighbor
charge-charge interaction leads to four separable kernels. If this
interaction is furthermore approximated by one kernel, associated with
the order parameter, Eq.(\ref{chi}) reduces to one scalar equation,
and one arrives at the models of Refs.\onlinecite{Maki1,Carbotte}.

The above formulas allow to calculate  density correlation functions
which can be expressed by $\rho'$ or $\rho$ defined in 
Eqs.(\ref{EF1})-(\ref{EF2}). The most general case may, however, 
involve two additional densities 
$\rho_0$ and $\rho'_0$
in Eqs.(\ref{EF1}) and (\ref{EF2}), respectively, where $E_0$ and $F_0$ 
are general functions of $\bf k$.
Going over to the Nambu representation $\hat{E}_0$ and $\hat{F}_0$ are
then linear combinations of the Pauli matrices and general functions of
${\bf k}$. Using the diagramatic rules and the effective Hamiltonian
one finds for the general susceptibility $\chi_{00}(q)$,
\begin{equation}
\chi_{00}(q) = \chi_{00}^{(0)}(q) + 
\sum_{\alpha\beta} \chi^{(0)}_{0\alpha}(q) \chi_{\alpha \beta}(q)
\chi^{(0)}_{\beta 0}(q).
\label{chiallgemein}
\end{equation}
$\chi^{(0)}_{00}(q)$ is given by Eq.(\ref{chi0}) where both $\alpha$
and $\beta$ have been replaced by $0$. Similarly, the expressions for 
$\chi^{(0)}_{0\beta}$ and $\chi^{(0)}_{\alpha 0}$ are obtained from
Eq.(\ref{chi0}) by replacing only $\alpha$ or $\beta$, respectively,
by $0$. $\chi_{\alpha\beta}$ is given by Eq.(\ref{chi}).

\section{Numerical results}
\subsection{D-wave susceptibility $\chi_{55}$}

\begin{figure}
\begin{center}
\setlength{\unitlength}{1cm}
\includegraphics[width=7cm,angle=0]{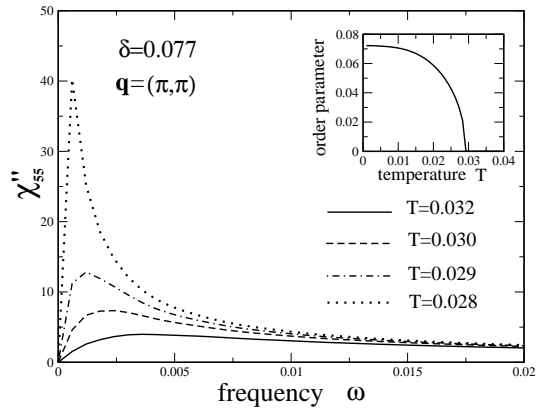}
\end{center}
\caption{
Negative imaginary part of $\chi_{55}$ of the $t$-$J$ model at large $N$
for $t'=-0.3$, $\eta =0.001$, ${\bf q} =(\pi,\pi)$, and 
doping $\delta=0.077$. Inset: $DDW$
order parameter $\Phi$ as a function of temperature $T$.
}
\label{fig1}
\end{figure}

From now on we will write $t =-|t|$ and use $|t|$ as the energy unit. 
Furthermore, the Heisenberg constant $J$ and the doping $\delta$ will 
be fixed to the values 0.3 and 0.077, respectively. The
mean-field value for the superconducting transition temperature is then
practically zero so that we deal with a pure DDW system. 
Fig.\ref{fig1} shows the negative imaginary part of the retarded 
susceptibility $\chi_{55}({\bf q},\omega)$, denoted by $\chi''_{55}$ in
the following,
as a function of $\omega$ in the normal state,
using $t'=-0.3$ and a small imaginary part $\eta = 0.001$. We have
chosen the component $\chi_{55}$ and the wave vector
${\bf q} = (\pi,\pi)$ because the transition
\noindent
to the DDW state occurs in this symmetry channel and near this wave vector. 
Decreasing the temperature from $T=0.032$ to $T=0.028$ shifts spectral weight
in the overdamped spectrum from large to small frequencies until
a sharp and intense peak appears very near to the transition point
\begin{figure}
\begin{center}
\setlength{\unitlength}{1cm}
\includegraphics[width=7cm,angle=0]{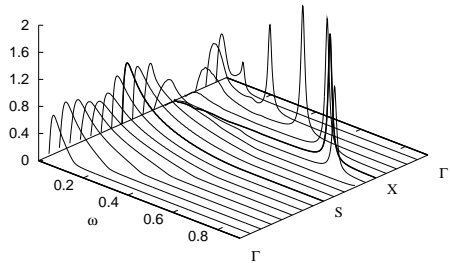}
\end{center}
\caption{
$\chi_{55}''$ in the normal state at $T=0.039$
calculated for $t'=-0.3$, $\delta=0.077$, and along the line 
$\Gamma$-$S$-$X$-$\Gamma$ in the Brillouin zone.
}
\label{fig2}
\end{figure}
\noindent
$T^\star \sim 0.028$. Decreasing further the temperature the spectrum
becomes again broad and moves to higher frequencies. The inset of 
Fig.\ref{fig1} shows the temperature dependence of the DDW order 
parameter $\Phi$ which exhibits the usual mean-field
behavior. 

Fig.\ref{fig2} shows a three-dimensional plot of 
$\chi''_{55}$ as a function of $\omega$ in the normal state at $T=0.039$. 
The momentum runs from the point $\Gamma=(0,0)$
to the points $S=(\pi,\pi)$ and $X=(0,\pi)$, and back to the point $\Gamma$. 
The curves for $\Gamma$, $S$ and $X$ are drawn as thick lines to 
facilitate the visiualization.  
According to Eq.(\ref{chi00}), and similar as in the case of the usual
charge susceptibility, $\chi_{55}$ exhibits a singular behavior at small 
frequencies and wave vectors in the limit $\eta \rightarrow 0$: 
Fixing the momentum to
${\bf q}=(0,0)$ $\chi_{55}(0,\omega)$ is zero for any finite frequency
$\omega$. Putting $\omega$ to zero and taking the limit ${\bf q} \rightarrow
0$ $\chi_{55}$ approaches a finite value.
This explains the absence of a thick line at $\Gamma$ in Fig.\ref{fig2}.
Moving the momentum from $\Gamma$ to $S$ the zero-frequency peak of
measure zero at $\Gamma$ aquires a finite width, i.e., a finite spectral 
weight, and at the same time peaks at a finite frequency. Both the
width and the peak increase first rapidly with momentum.
The peak position passes then through a maximum at $\omega \sim 0.04$, starts 
to decrease, and reaches a minimum at $S$ signalling the transition
to the DDW state at lower temperatures. The width of the peak as well
as the extension and intensity of the slowly decaying structureless background 
increases up to the point $S$. Moving along the direction 
$S$-$X$-$\Gamma$ the low-frequency peak looses rapidly
spectral weight, vanishes
practically at $X$, and recovers somewhat towards $\Gamma$. At the same time
a well-pronounced, strong high-frequency peak develops, its frequency 
decreases monotonically along the above path, and its intensity shows 
a maximum near the point $X$.
The dispersion of this high-frequency peak is $\sim \sin{k_x}$ between
$\Gamma$ and $X$, i.e., it is identical with the sound wave due to usual
density fluctuations\cite{Gehlhoff}. This means that d-wave density 
fluctuations,
described by $\chi_{55}''$, strongly couple to usual density fluctuations
along the path $S$-$X$-$\Gamma$ whereas such a coupling is forbidden by
symmetry along $\Gamma$-$S$.     
\begin{figure}
\begin{center}
\setlength{\unitlength}{1cm}
\includegraphics[width=7cm,angle=0]{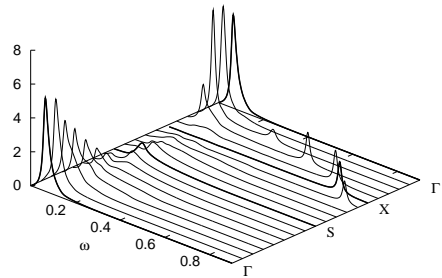}
\end{center}
\caption 
{$\chi_{55}''$ at T=0.
calculated for $t'=-0.3$, $\delta=0.077$, and along the lines 
$\Gamma$-$S$-$X$-$\Gamma$ in the Brillouin zone.
}
\label{fig3}
\end{figure}

Fig.\ref{fig3} shows the same 3d plot as 
Fig.\ref{fig2} but the calculation is now performed at T=0 in the DDW 
state. Differences in the two figures thus have to be ascribed 
to the formation of the DDW. Taking the different scales in the two figures 
into account one sees that the intensity of the high-frequency peak due 
to usual density fluctuations did not change much. Big changes, however, occur
at low frequencies. $\chi_{55}''$ no longer vanishes at finite
frequencies at the $\Gamma$-point but rather shows there a well-pronounced
and strong peak. Moving away from the $\Gamma$ point this peak looses
rapidly intensity whereas its frequency does not change much.

\begin{figure}
\begin{center}
\setlength{\unitlength}{1cm}
\includegraphics[width=7cm,angle=0]{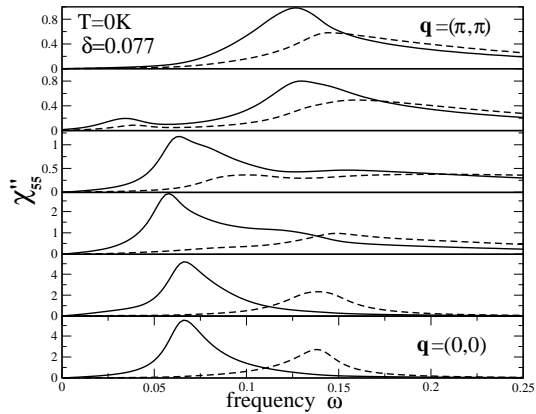}
\end{center}
\caption 
{$\chi_{55}''$ (solid lines) and
$\chi_{55}^{(0)''}$ (dashed lines) for momenta along the diagonal
${\bf q}=(q,q)$.
}
\label{fig4}
\end{figure}

To understand the origin and the properties of the low-frequency peak in
more detail we have plotted in Fig.\ref{fig4} $\chi_{55}''$ and 
${\chi_{55}^{(0)}}''$ as a function of frequency for several $\bf q$ values 
between the $\Gamma$ and $S$ point. The dashed line in the upper diagram 
represents ${\chi_{55}^{(0)}}''$ at the wave vector
${\bf q}=(\pi,\pi)$. It contains only interband transitions across
the DDW gap, mainly between the points $X$ and $Y$ and between the hot spots
on the boundaries of the RBZ. The intraband contribution is zero by symmetry
at $S$. Since $\Phi$ is about 0.07 the interband transitions lead to a broad
peak somewhat higher than twice the DDW gap. The solid curve in the upper 
diagram of Fig.\ref{fig4} represents $\chi_{55}''$ at $S$. It shows a much
more pronounced peak than the dashed curve which originates from the
denominator in Eq.(\ref{chi55}), i.e., which represents a collective
excitation. It corresponds to the amplitude mode of the DDW,
where the order parameter $\Phi$ is modulated without changing its
d-wave symmetry. Its frequency is somewhat smaller than $2\Phi$ which is
expected for a d-wave state. For an isotropic s-wave ground state the two 
frequencies would be exactly the same.   
\noindent

The lowest panel in Fig.\ref{fig4} shows the same susceptibilities at
the $\Gamma$ point. ${\chi_{55}^{(0)}}''$ again contains no intraband
but only interband contributions consisting of vertical transitions
which probe directly the DDW gap. As a result, ${\chi_{55}^{(0)}}''$
exhibits a well-defined peak at $2\Phi$ and a long-tail towards smaller
frequencies due to the momentum dependence of the order parameter.
The solid curve, representing $\chi_{55}''$, exhibits a well-pronounced
peak much below $2\Phi$. It corresponds to a bound state inside the
d-wave gap created by multiple scattering of the excitations across
the gap due to the Heisenberg interaction. Due to its large binding energy
the density of states inside the gap is rather small at the peak position
so that the width of the peak is substantialy smaller than that of the
amplitude mode at the momentum
$(\pi,\pi)$. Previously\cite{Zeyher4}, we have associated this peak with the 
$B_{1g}$ peak seen in Raman scattering in high-T$_c$ superconductors.
Note that there is roughly a factor 5 difference in the scales  
of the top and bottom panels in Fig.\ref{fig4}. It means that the amplitude
mode of the DDW order parameter has a much larger spectral weight and also
smaller width at the $\Gamma$ than at the $S$ point. 

For momenta $\bf q$ between $\Gamma$ and $S$ both
intra- and interband transitions contribute to ${\chi_{55}^{(0)}}''$.
The intraband part, reflecting the residual Fermi arcs in the DDW state,
turns out to be in general substantially smaller than the interband part
and consists of a broad peak which disperses roughly as $\sin(q)$
and which looses its intensity when approaching $\Gamma$ or $S$. 
So only the low-frequency tails and the structure around
the frequency 0.04 in the second panel from above can be attributed to 
intraband scattering, the remaining spectral weight is due to interband
scattering. Going from $\Gamma$ to $S$ the solid lines show the 
dispersion of the amplitude mode which is in general well below the main 
spectral peak in ${\chi_{55}^{(0)}}''$, representing a resonance mode
with a frequency and width determined mainly by the denominator in
Eq.(\ref{chi55}). From the different scales used in each panel it is
evident that the intensity of the amplitude mode strongly decays
away from the $\Gamma$ point in agreement with Fig.\ref{fig3}.  

\subsection{Non-resonant inelastic X-ray scattering}

Specifying the polarization vectors of the incident and scattered light
and using Eqs.(\ref{eps}) and (\ref{dispersion}) for the electron 
dispersion the cross
section for inelastic X-ray scattering is determined by $\chi_X''$ 
and $\chi_X$ is obtained as a linear combination of the susceptibility matrix
elements $\chi_{\alpha \beta}$. As an example, consider the case
${\bf e}^i =(1,1)/\sqrt{2}, {\bf e}^s =(1,-1)/\sqrt{2}$,
which yields for ${\bf q} = 0$ 
the $B_{1g}$ component of Raman scattering. Choosing the symmetry
direction ${\bf q}=(q,q)$, we find,   
\begin{eqnarray}
\chi_X ({\bf q},i\omega_n) = {2 t_{eff}^2 \over {J}}\Bigl(
\cos{^2}({{q}\over 2}) \chi_{55}({\bf q},i\omega_n)  \nonumber  \\
 + \sin{^2}({q\over 2}) \chi_{66}({\bf q},i\omega_n)  
-\sin(q) \chi_{56}({\bf q},i\omega_n)\Bigr),
\label{chiX}
\end{eqnarray}
\begin{figure}
\begin{center}
\setlength{\unitlength}{1cm}
\includegraphics[width=7cm,angle=0]{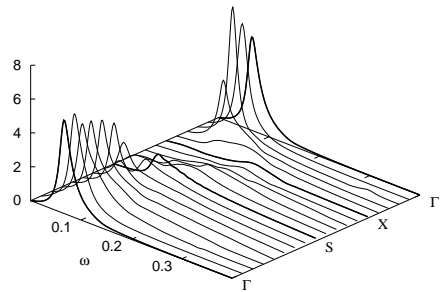}
\end{center}
\caption 
{D-wave cross section for nonresonant, inelastic $X$-ray scattering
for the $t$-$J$ model at large $N$ in the DDW state at $T=0$. The momentum
varies along the points $\Gamma$-$S$-$X$-$\Gamma$ in the Brillouin zone.   
}
\label{fig5}
\end{figure}
with the effective nearest-neighbor hopping element $t_{eff}$ determined 
by the large $N$ dispersion of Eq.(\ref{eps}). 
The second-nearest hopping element $t'$ drops
out in the considered symmetry. 
In Fig.\ref{fig5} we have plotted 
$\chi_X$ without this prefactor along the points $\Gamma$-$S$-$X$-$\Gamma$ 
generalizing
Eq.(\ref{chiX}) to an arbitrary ${\bf k}$ point in the Brillouin zone. 
One feature of $\chi_X''$ is that 
the high-frequency side bands due to
usual density fluctuations are practically absent. This may be explained
by the fact that  
the various contributions to $\chi_X ''$ have both signs  
causing large cancellation effects. 
Another difference between Figs.\ref{fig3} and \ref{fig5}
occurs around the point $S$ where
additional structure is seen in $\chi_X''$ due to the component $\chi_{66}$.

\subsection{Inelastic neutron scattering} 

Using the symmetry-adapted functions Eq.(\ref{basis}) the expression for
$\gamma$, Eq.(\ref{gamma}), can be written as,
\begin{equation}
\gamma({\bf k},{\bf q}) = {{8\pi ie \mu_0 t_{eff}}\over{\hbar c|{\bf q}|}}
\sum_{\beta=3}^6 B_\beta({\bf q}) \gamma_\beta({\bf k}),
\label{gam2}
\end{equation}
with
\begin{equation}
B_{3,5} = {\alpha_x \over q_x}(1-cos q_x) \pm {\alpha_y\over q_y}
(1-cos q_y),
\label{B1}
\end{equation}
\begin{equation}
B_{4,6} = {\alpha_x\over q_x} sin q_x \pm {\alpha_y\over q_y} sin q_y,
\label{B2}
\end{equation}
where the subscripts $3,4$ refer to the $+$ and $5,6$ to the $-$ signs, 
respectively.
The charge correlation function, describing neutron scattering, is
given by,
\begin{equation}
\chi_N({\bf q},\omega) = {1\over{2J}}\Bigl( {{8\pi e \mu_0 t_{eff}}\over 
{c|{\bf q}|}}
\Bigr)^2 \sum_{\beta,\delta} B_\beta({\bf q}) B_\delta ({\bf q})
\chi_{\beta \delta}({\bf q},\omega).
\label{chiN}
\end{equation}
\vspace{.15cm}
\begin{figure}
\begin{center}
\setlength{\unitlength}{1cm}
\includegraphics[width=7cm,angle=0]{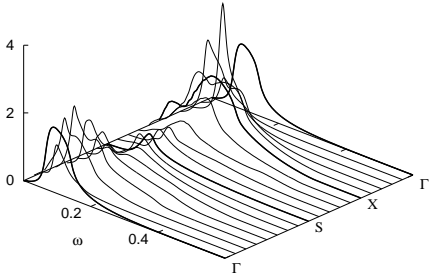}
\end{center}
\caption 
{$\chi_N''$ describing neutron scattering as a function of $\omega$
along the line $\Gamma$-$S$-$X$-$\Gamma$ in the Brillouin zone 
in the DDW state at $T=0$ using $t'=-0.3$, $\delta = 0.077$,
$q_z=0.5$, and $\eta = 0.05$. 
}
\label{fig6}
\end{figure}

Fig. \ref{fig6} shows $\chi_N''({\bf q},\omega)$ as a function of
$\omega$ for momenta along the symmetry line $\Gamma$-S-X-$\Gamma$ in the
Brillouin zone for the fixed momentuma transfer $q_z=0.5$. The 
spin of the neutrons was assumed to be polarized along the $z$-axis.
The spectra show essentially one more or less 
well-defined peak which disperses roughly between the frequencies
$0.1 - 0.2$. Similarly as in Fig.\ref{fig4} this peak describes a mainly 
collective excitation of the system related to a bound state 
inside the DDW gap. In spite of the rather large momentum
transfer $q_z =0.5$ the peak intensity increases
substantially towards the point $\Gamma$. This increase becomes more and more 
pronounced if $q_z$ is further lowered 
until it diverges like $1/|{\bf q}|^2$ in the limit $q_z \rightarrow 0$.

An important special case of Eq.(\ref{chiN}) is scattering by unpolarized
neutrons. Averaging over all spin directions for the neutrons yields
the charge correlation function $\chi_{N,u}$ for scattering with 
unpolarized neutrons,
\begin{equation}
\chi_{N,u}({\bf q},\omega) = \sum_{\alpha\beta} F_{\alpha\beta}({\bf q})
\chi_{\alpha \beta}({\bf q},\omega),
\label{chiNu1}
\end{equation}
with
\begin{equation}
F_{\alpha\beta} = {1\over{2J}}\Bigl( {{8\pi e \mu_0 t_{eff}}\over 
{c|{\bf q}|}}\Bigr)^2 
<B_\alpha({\bf q}) B_\beta ({\bf q})>_{av},
\label{chiNu2}
\end{equation}
where $<...>_{av}$ denotes an average over all directions of the moment
of the neutron. We obtain,
\begin{eqnarray}
<B_\alpha({\bf q}) B_\beta ({\bf q})>_{av} = 
{1\over {|{\bf q}|^2}}[(q_y^2+q_z^2)f_\alpha f_\beta \nonumber\\
-q_xq_y(f_\alpha g_\beta
+g_\alpha f_\beta) +(q_x^2+q_z^2)g_\alpha g_\beta],
\label{fg}
\end{eqnarray}
with $f_3 = f_5 = (1-\cos{q_x})/q_x$, $g_3 = -g_5 = (1-\cos{q_y})/q_y$, 
$f_4 = f_6 = \sin{q_x}/q_x$, $g_4 = -g_6 = \sin{q_y}/q_y$. 
$\chi_{N,u}({\bf q},\omega)$ diverges at small momentum transfers in the
DDW state. To extract its singular part one may take the limit
${\bf q} \rightarrow 0$ in the $f$, $g$, and $\chi$ functions. One obtains
then,
\begin{equation}
\chi_{N,u} \sim ({1\over{|{\bf q}|^2}}+ {q_z^2 \over{|{\bf q}|^4}})
\chi_{66}(0,i\omega_n).
\label{sing}
\end{equation}

In the normal state $\chi_{66}$ vanishes at ${\bf q}=0$ for any finite 
frequency as can be seen from the explicit expression Eq.(\ref{chi00}). 
In contrast to that
it is finite at ${\bf q}=0$ in the DDW state describing a continuum of
particle-hole excitations with zero total momentum across the gap.
As a result $\chi_{N,u}$ diverges quadratically in the momentum ${\bf q}_{\|}$
parallel to the layers for $q_z=0$. This divergence is caused by the
long-range part of the interaction between the magnetic moment of the neutron
and the electrons in the layers. It is thus a real effect causing a singular
forward scattering contribution in the cross section. 
Integrating $\chi_{N,u}''$ over ${\bf q}$ diverges logarithmically in a 
strictly 
2d description at $q_z=0$ whereas it remains finite in three dimensions.
\vspace{0.cm}
\begin{figure}
\begin{center}
\setlength{\unitlength}{1cm}
\includegraphics[width=7cm,angle=0]{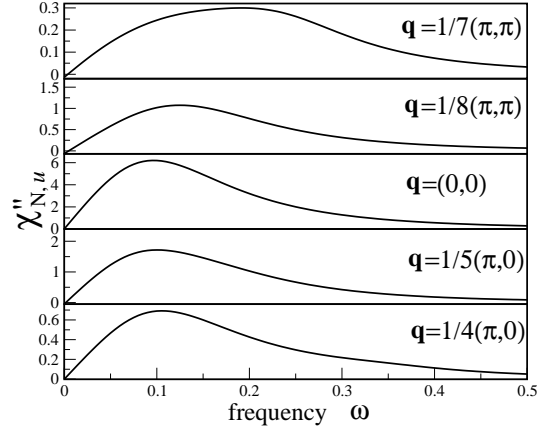}
\end{center}
\caption{ 
$\chi_{N,u}''$ describing scattering with unpolarized neutrons 
as a function of $\omega$
for momenta around the point $\Gamma$ 
in the DDW state at $T=0$ using $t'=-0.3$, $\delta = 0.077$ and 
$q_z=0.5$. 
}
\label{fig7}
\end{figure}
The large enhancement of $\chi_{N,u}''$ near the point $\Gamma$
is illustrated in Fig.\ref{fig7}. The curve at $\Gamma$ shows
a peak at $\omega \sim 0.1$ well inside the gap of $2\Delta_0 \sim 0.14$,
which disperses to higher frequencies moving away from $\Gamma$, especially,
along the diagonal. More dramatic is, however, the rapid decay
of the intensity of the peak away from $\Gamma$ which  can be seen  
from the different scales used in plotting the various panels.

Another special case of Eq.(\ref{chiN}) is obtained for an arbitrary 
polarization for the neutron moment and a vanishing momentum transfer 
in the $z$ direction, i.e., for ${\bf q} = (q_x,q_y,0)$. The functions
$B_i$ assume in this case the form for $i=3,4$,
\begin{eqnarray}
B_i = {\mu_z\over{\mu_0 |{\bf q}|}} (-q_y f_i +q_x g_i), \nonumber \\
B_{i+2} =  {\mu_z\over{\mu_0|{\bf q}|}} (-q_y f_i - q_x g_i).
\label{BB}
\end{eqnarray}
The polarization dependence of $\chi_N$ is simply $\cos{^2}\theta$ where
$\theta$ is the angle between the neutron spin and the normal to the 
layers. $\chi_N$ assumes its largest value if the spin is perpendicular and
is zero if the spin is parallel to the layers. In this case $\chi_N$
diverges as $ 1/|{\bf q}|^2$ for all polarizations of the neutron and the
singular contribution is isotropic in the $q_z=0$ plane. For $q_z \neq 0$
the dependencies of $\chi_N$ on the polarization and the momentum no longer 
factorize but can be worked out for the most general case using
Eq.(\ref{chiN}).
 
\subsection{Comparison of scattering efficiencies from spin and 
orbital fluctuations}

As discussed in the previous sections the total magnetic neutron cross
section contains, in addition to the dominating spin flip contribution,
an orbital part. Above and not too far away from $T^\star$ to the 
DDW state the scattering from fluctuating orbital moments
produced by circulating currents around a plaquette are mainly concentrated
around the wave vector ${\bf Q} = (\pi,\pi)$. Below $T^\star$ there is an 
elastic
Bragg component at wave vector ${\bf Q}$ and a fluctuating part 
throughout the Brillouin zone. Though magnetic scattering from spin flip
and orbital scattering have many features in common there are
three main differences: a) In spin scattering the form factor reflects
the electron density distribution within an atom whereas in orbital
scattering the form factor is related to the plaquette geometry which gives 
rise to the trigonometric functions in the functions $B_\beta$; 
b) The long-range dipole-dipole interaction between neutrons and the 
electrons dominates in
orbital scattering yielding a singular forward scattering peak 
below $T_c$ whereas the interaction 
between neutrons and electrons for spin scattering reduces to a local one;
c) The intensity of orbital scattering is substantially smaller than that for 
spin scattering. For instance, a ratio of 1:70 has been obtained in 
Ref.\onlinecite{Hsu} for Bragg scattering. In the following we present
results on frequency and in-plane momentum integrated scattering 
probabilities in the two cases. In this way the anomalous forward scattering 
contribution in orbital scattering can be taken into account
in this comparison. We also recalculate the intensity for Bragg scattering
and find it much smaller than the value calculated in Ref.\onlinecite{Hsu}.

The expression $2 \chi_{N,u}''({\bf q},\omega)(1+n(\omega))$ 
describes the scattering
probability of an unpolarized neutron with momentum and frequency
transfers ${\bf q}$ and $\omega$, respectively, from orbital fluctuations.
Integrating over frequency and the in-plane momentum ${\bf q}_{\|}$
the integrated scattering probability $P_{orb}$ becomes, using 
Eq.(\ref{chiNu1}) and reinserting $a$ and $\hbar$ for clarity,
\begin{equation}
P_{orb}(q_z) = P_{orb}^{Bragg}(q_z) + P_{orb}^{dyn}(q_z),
\label{Wtotal}
\end{equation}
\begin{equation}
P_{orb}^{Bragg}(q_z) = 2F_{55}({\bf Q},q_z) 
\langle \rho_5'({\bf Q}) \rangle
\langle \rho_5^\dagger({\bf Q}) \rangle,
\label{WBragg}
\end{equation} 
\begin{eqnarray}
P_{orb}^{dyn}(q_z) &=& {2\over {N_c^{2/3}}} \sum_{{\bf q}_{\|}}
\int^\infty_{-\infty} d \omega (1+n(\omega)) 
\tilde{\chi}_{N,u}''({\bf q},\omega) \nonumber \\
&=& {2\over N_c^{2/3}} \sum_{{\bf q}_{||},\alpha,\beta}
F_{\alpha\beta}({\bf q}) \langle \tilde{\rho}_\alpha'({\bf q}_{\|}) 
\tilde{\rho}_\beta^\dagger({\bf q}_{\|}) \rangle .
\label{Worb}
\end{eqnarray}
In Eq.(\ref{Wtotal}) $P_{orb}$ has been splitted into a static 
Bragg contribution $P_{orb}^{Bragg}$ and a
dynamic part $P_{orb}^{dyn}$. Correspondingly, the tilde in 
Eq.(\ref{Worb}) indicates that only the fluctuating part of the densities
should be used in the correlation functions.

Similar considerations apply to unpolarized neutron scattering from spin 
fluctuations.
The relevant susceptibility in the paramagnetic state is\cite{Kittel}, 
\begin{equation}
{{\mbox{\boldmath$\mu$}\unboldmath ^2}\over 6}
\Bigl( {{8\pi e F({\bf q})}\over{mca^3}}\Bigr)^2 \chi_{zz}({\bf q},\omega).
\label{spinchi}
\end{equation}
$F({\bf q})$ is the atomic structure factor and $\chi_{zz}({\bf q},\omega)$
the zz component of the electronic spin susceptibility. Using the 
fluctuation-dissipation
theorem, approximating $F({\bf q})$ by $F(0)=1$, and assuming no magnetic
coupling between the layers, the total scattering
probability $P_s$ by spin fluctuations becomes\cite{Greco},
\begin{equation}
P_s =  {{\mbox{\boldmath$\mu$}\unboldmath ^2}\over{12}}
\Bigl( {{8\pi e }\over{mca^3}}\Bigr)^2 \pi(1-\delta).
\label{Ws}
\end{equation}
Most interesting are the ratios of integrated orbital to spin scattering,
i.e.,  $R^{Bragg} = P_{orb}^{Bragg}/P_s$ and 
$R^{dyn} = P_{orb}^{dyn}/P_s$. Using the above results one obtains,
\begin{eqnarray}
R^{dyn}(q_z) &=& A {1\over N_c^{2/3}} \sum_{{\bf q}_{\|},\alpha,\beta}
{{\langle B_\alpha({\bf q}) B_\beta({\bf q}) \rangle}_{av}\over
{(a{\bf q})^2}} \nonumber \\
\cdot \int^{\infty}_{-\infty} &d(\hbar)\omega& (1+n(\omega))
\chi''_{\alpha\beta}({\bf q}/|t|,
\omega),
\label{Rdyn}
\end{eqnarray}
with the dimensionless constant,
\begin{equation}
A={4\over{\pi (1-\delta)(J/|t|)}} \Bigl( {t_{eff} 
\over{\hbar^2/(m a^2)}}\Bigr)^2,
\label{A}
\end{equation}
and
\begin{equation}
R^{Bragg}(q_z) = A {{\langle (B_5({\bf Q}))^2\rangle_{av}}\over 
{a^2({\bf Q}^2 +q_z^2)}} \langle \rho_5'({\bf Q}) \rangle
\langle \rho_5^\dagger({\bf Q}) \rangle.
\label{RBragg}
\end{equation}

\begin{figure}
\begin{center}
\setlength{\unitlength}{1cm}
\includegraphics[width=7cm,angle=0]{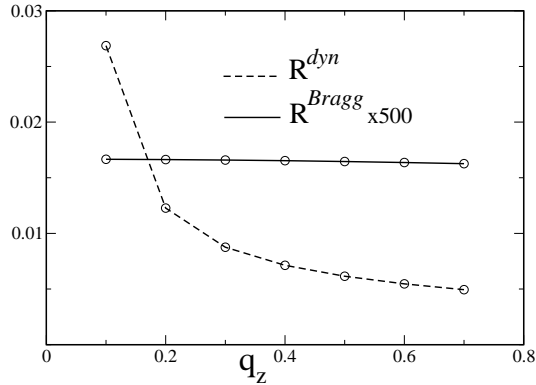}
\end{center}
\caption{
Ratios $R^{Bragg}$ and $R^{dyn}$ for elastic and inelastic orbital to spin 
scattering, respectively.
}
\label{fig8}
\end{figure}

Using $|t|=0.5\; eV$ we obtain from Eq.(\ref{eps}) $|t_{eff}| 
\sim 0.07\;eV$, which
yields togther with $a=3.856$ \AA \; 
$A \sim 0.09$ for our
doping $\delta = 0.077$. Comparison of Eq.(\ref{Heff1}) with Eq.(\ref{H0})
gives 
\begin{equation}
\langle \rho_5'({\bf Q}) \rangle \langle \rho^\dagger_5({\bf Q}) \rangle
= { {\Phi}^2\over{2J}} \sim 0.082,
\label{num}
\end{equation}
for $T=0$ and $\delta = 0.077$. Furthermore, from Eq.(\ref{fg}) we get
\begin{equation}
\langle (B_5({\bf Q}))^2 \rangle_{av} = 8/\pi^2,
\end{equation} 
which yields 
\begin{equation}
R^{Bragg}(0) = 3 \cdot 10^{-5}.
\end{equation} 
Unlike in the case of spin scattering the momentum and frequency sums
in Eq.(\ref{Rdyn}) are non-trivial and we have carried them out by numerical 
methods.

Fig. 8 shows the results for $R^{dyn}$ and $R^{Bragg}$ as a function
of $q_z$ at zero temperature. As expected $R^{dyn}$ increases strongly
at small $q_z$ due to its logarithmic singularity at $q_z=0$. Integrating
$R^{dyn}$ also over $q_z$ yields a value of about $10^{-2}$ which is more
than 2 orders of magnitude larger than $R^{Bragg}$. The reason for this
large difference is due to the fact that in the momentum integration
in $R^{dyn}$ mainly the small momenta contribute where $1/{\bf q}^2$
is larger than 1. In contrast to that, $R^{Bragg}$ probes the integrand 
of $R^{dyn}$ at the large value $|{\bf Q}|$ where  $1/{\bf q}^2$
is very small. The extreme small value $R^{Bragg} \sim 10^{-5}$ 
casts doubts on the identification
of the observed Bragg peak at $\bf Q$ with the Bragg peak due a
DDW\cite{Mook}. On the other hand,
if a DDW state is realized in some material we think that the resulting
strong forward scattering peak in neutron scattering would be 
observable. According to our predictions it would exhaust roughly
3 per cent of the sum rule for spin scattering and thus be of the
same order of magnitude as 
the recently detected elastic scattering from weak magnetic moments
directed at least partially perpendicular to the layers\cite{Bourges} in 
high-$T_c$ superconductors.

\section{Conclusions}

The excitation spectrum of the $t$-$J$ model at large $N$ was studied for 
dopings 
where a DDW forms below a transition temperature $T^\star$. The density 
response includes ``conventional'' local density 
fluctuations, characterized by the energy scale $t$, where practically all
spectral weight is concentrated in a collective sound wave with  
approximately a sinusoidal dispersion. In addition, and this was the main
topic of this investigation, there is a 
``unconventional'' density response from orbital fluctuations caused by 
fluctuating
circulating currents above $T^\star$ and fluctuations around a staggered 
flux phase below $T^\star$. The corresponding spectra are again mainly
collective, their energy scale is $J$ or a fraction of it, and they can
be viewed as order parameter modes of the DDW. At small
momentum transfers the spectra are dominated by one well-defined peak
corresponding to amplitude fluctuations of the DDW.
For larger momentum transfers,  
unconventional density fluctuations tend to become broad and their intensity
suppressed. This is especially true near the point X where practically
no weight is left for orbital fluctuations at low frequencies, 
but instead a rather sharp peak due to sound waves appear at large
frequencies reflecting the coupling of the two kinds of density fluctuations.

In principle the above orbital fluctuations show up in the Raman,
X-ray and neutron scattering spectra. The predicted scattering intensities 
are, however, rather small. The reason for this is that, on the one hand,
a rather small effective nearest-neighbor hopping element $t_{eff}$ is
needed to provide a sufficiently large density of states and the instability
to the DDW, on the other hand, the square of $t_{eff}$ appears as a prefactor
in the various cross sections due to the Peierls substitution. As a result,
the $\bf q$ integrated inelastic cross section is only a few per cents
of that for spin scattering and the elastic scattering from the Bragg
peak is even much weaker. From our calculations it is also evident that a
DDW state in the unconstrained, weak-coupling case would have very similar 
properties as in the $t$-$J$ model. In particular, the curves for
$R^{dyn}$ and $R^{Bragg}$ would closely resemble those in Fig.\ref{fig8}.
An experimental verification of a DDW state via inelastic
neutrons seems to be not unfeasible because a good deal of the scattering
intensity is concentrated in a well-pronounced, inelastic peak somewhat
smaller than the DDW gap and in small region around the origin in $\bf k$
space.
     
{\bf{Acknowledgements}} The authors are grateful to M. Geisler and T. Enss for 
help with the figures, H. Yamase for useful discussions, and D. Manske for 
a critical reading of the manuscript.

%\end{multicols}
\end{document}